\documentclass[]{spie}  

\usepackage{amsmath,amsfonts,amssymb}
\usepackage{bbm}
\usepackage{bbold}
\usepackage{subfig}

\usepackage{booktabs}
\usepackage{graphicx}
\usepackage{caption}
\usepackage[colorlinks=true, allcolors=blue]{hyperref}

\title{Benchmarking Image Transformers for Prostate Cancer Detection from Ultrasound Data}

\author[1]{Mohamed Harmanani}
\author[1]{Paul F. R. Wilson}
\author[2]{Fahimeh Fooladgar}
\author[1]{Amoon Jamzad}
\author[1]{\\Mahdi Gilany}
\author[2]{Minh Nguyen Nhat To}
\author[3]{Brian Wodlinger}
\author[2]{Purang Abolmaesumi}
\author[1]{\\Parvin Mousavi}
\affil[1]{Queen's University, Kingston, Canada}
\affil[2]{University of British Columbia, Vancouver, Canada}
\affil[3]{Exact Imaging, Markham, Canada}

\authorinfo{Send correspondance to: 22mh5@queensu.ca}

\begin{document} 
\maketitle

\begin{abstract}
\textbf{PURPOSE:} Deep learning methods for classifying prostate cancer in ultrasound images typically employ convolutional networks to detect cancer in small regions of interest (ROI) along a needle trace region. Recently, multi-scale approaches have combined the context awareness of Transformers with a convolutional feature extractor to detect cancer from multiple ROIs. In this work, we present a detailed study of several Image Transformer architectures for both ROI-scale and multi-scale classification, and a comparison of the performance of CNNs and Transformers for ultrasound-based prostate cancer classification. \textbf{METHODS:} We use a dataset of 6607 prostate biopsy cores extracted from 693 patients at 5 distinct clinical centers. We evaluate 3 vision transformers on ROI-scale cancer classification then use the strongest model to tune a multi-scale classifier using multi-objective learning. We compare our results in both settings to a baseline convolutional architecture typically used in computer vision tasks. We evaluate all our models using nested $k$-fold cross-validation. \textbf{RESULTS}: Our core-wise multi-objective model achieves a 77.9\% AUROC, a sensitivity of 75.9\%, and a specificity of 66.3\%, a
considerable improvement over the baseline. \textbf{CONCLUSION}: We conclude that the combined use of Image Transformers and multi-objective learning has the potential to improve performance in prostate cancer classification from ultrasound. 
\end{abstract}


\section{PURPOSE}
\label{sec:intro}  
Early and accurate diagnosis of prostate cancer (PCa) is crucial in order to improve the chances of successful treatment. The standard method used to diagnose PCa is the histopathological annotation of biopsy tissue retrieved from the patient under the guidance of Trans-rectal ultrasound (TRUS). Because of the low sensitivity of conventional ultrasound in identifying prostate lesions~\cite{ahmed2017diagnostic}, TRUS-guided biopsy is usually systematic in nature: a number of biopsy cores are sampled from different locations in the prostate. This is in contrast to targeted biopsy, where tissue is sampled from a specific suspicious tissue location. Systematic biopsy carries significant risks of adverse effects due to the large number of biopsy samples that need to be retrieved. As such, improving the performance of targeted biopsy has the potential to decrease the likelihood of biopsy-related risks and complications. 

\noindent Micro-ultrasound is a newly developed technology that allows visualization of tissue microstructures at much higher resolutions than conventional ultrasound. As such, this imaging modality is a prime candidate for training deep learning models to detect prostate cancer in ultrasound images. Typically, deep learning is used during targeted biopsy to classify small regions of interest (ROI) across a needle trace region\cite{wilson2022self}. This approach has seen some measure of success, but still struggles with a number of issues. For instance, ROI-scale PCa detection suffers from weak labelling: ground-truth histopathology labels describe tissue properties of the entire biopsy core, and ROI labels are only an approximation of the true distribution of cancer in the core. Moreover, ROI-scale models do not consider the broader contextual information encoded in multiple overlapping patches as clinicians typically~do. 

\noindent Multi-scale methods to PCa detection from ultrasound have recently been proposed as a solution to these problems, and have been shown to outperform ROI-scale classifiers\cite{gilany2023trusformer}. Recent work has shown that using a Transformer model to leverage contextual information in multiple patches belonging to the same core can increase the performance of deep learning-based PCa detection. In this work, we explore the effectiveness of Vision Transformers as feature extractors in both ROI-scale and multi-scale contexts. We perform a detailed study of several architectures in the hopes of improving the performance of multi-scale PCa detection. Furthermore, we introduce a novel learning objective that takes advantage of both core-scale and ROI-scale predictions to improve multi-scale Transformer models for PCa detection.

\section{MATERIALS AND METHODS}

\subsection{Data Collection \& Processing}
We use data collected from 693 patients who underwent prostate biopsy in five centers under the guidance of Trans-rectal ultrasound (TRUS). The biopsy cores are extracted using the ExactVu micro-ultrasound system~\cite{rohrbach2018high}. Raw Radio Frequency (RF) ultrasound images of the tissue are saved immediately before the biopsy gun is fired, and the needle trace region is approximately determined using the angle and depth of the penetration. The extracted biopsy cores are then analyzed histopathologically to determine the Gleason score and the approximate percentage of cancer present. We have 6607 total cores, with 86.7\% of the dataset being non-cancerous. To mitigate the imbalance of the labels, we undersample the benign cores during training in order to ensure the dataset has an equal amount of benign and cancerous cores. Using the known angle and depth of the needle's penetration, we determine a rectangular trace region. 

\noindent Regions of interest (ROIs) of dimension $5\times 5$ are extracted from the areas where the needle trace region overlaps with the prostate mask and are labelled 0 (benign) or 1 (cancer). Finally, each ROI is reshaped to $256\times256$ using linear interpolation, instance-normalized by computing its mean and standard deviation, and rescaled to the range (0, 1). To build our dataset, we extract 55 patches along the needle region of each core and compile them into training, validation, and test sets. We make sure to keep cores and patches from the same patient in the same set to avoid any risk of data leakage. 

\subsection{Self-supervised Pre-training}
We use Variance-Invariance-Covariance Regularization\cite{bardes2021vicreg} (VICReg) to pre-train our models. VICReg works in the following manner: given an ultrasound image $x$, two stochastic data augmentations $t_1$ and $t_2$ are applied to the image to yield two distinct views of $x$, denoted $x_1$ and $x_2$. For example, the image may be randomly rotated, scaled, cropped, or distorted~\cite{bardes2021vicreg}. Afterwards, an encoder model is used to extract vector representations $h_1$ and $h_2$ from each view, which are then projected into a latent space by a MLP network, where the following self-supervised loss function is applied:
\begin{align*}
    \mathcal{L}_\text{VICReg}(z_1, z_2) &= \lambda s(z_1,z_2) + \mu[v(z_1) + v(z_2)] + \nu[(c(z_1) + c(z_2)],
\end{align*}
where $z_1$ and $z_2$ are the projections of $h_1$ and $h_2$, and $\lambda$, $\mu$, $\nu$ are tunable hyperparameters. Following prior work, we use $\lambda=25, \mu=25, \nu=1$. \cite{wilson2022self, gilany2023trusformer, bardes2021vicreg}. $s, v$, and $c$ designate the invariance, variance, and covariance loss functions respectively. The invariance loss is given by the mean squared error loss (MSE), the variance loss maintains the variance of features across batches, and the covariance loss minimizes feature redundancy through cross-correlations projected features\cite{wilson2022self, bardes2021vicreg}.  

\subsection{Supervised Finetuning}
\noindent We compare several transformer architectures on the task of cancer detection on a single ROI. To that end, we use a standard Vision Transformer\cite{dosovitskiy2020image} (ViT) architecture, a Compact Convolutional Transformer\cite{hassani2021escaping} (CCT), and a Pyramid Vision Transformer\cite{wang2021pyramid} (PvT). We choose a modified ResNet18~\cite{he2016deep} as our ROI-scale baseline, with only one sequence of convolutions and batch normalization in each residual block. This reduction in the number of parameters mitigates overfitting and is associated with an increase in performance\cite{wilson2022self}.

\begin{figure}[ht!]
    \centering
    \includegraphics[width=0.92\columnwidth]{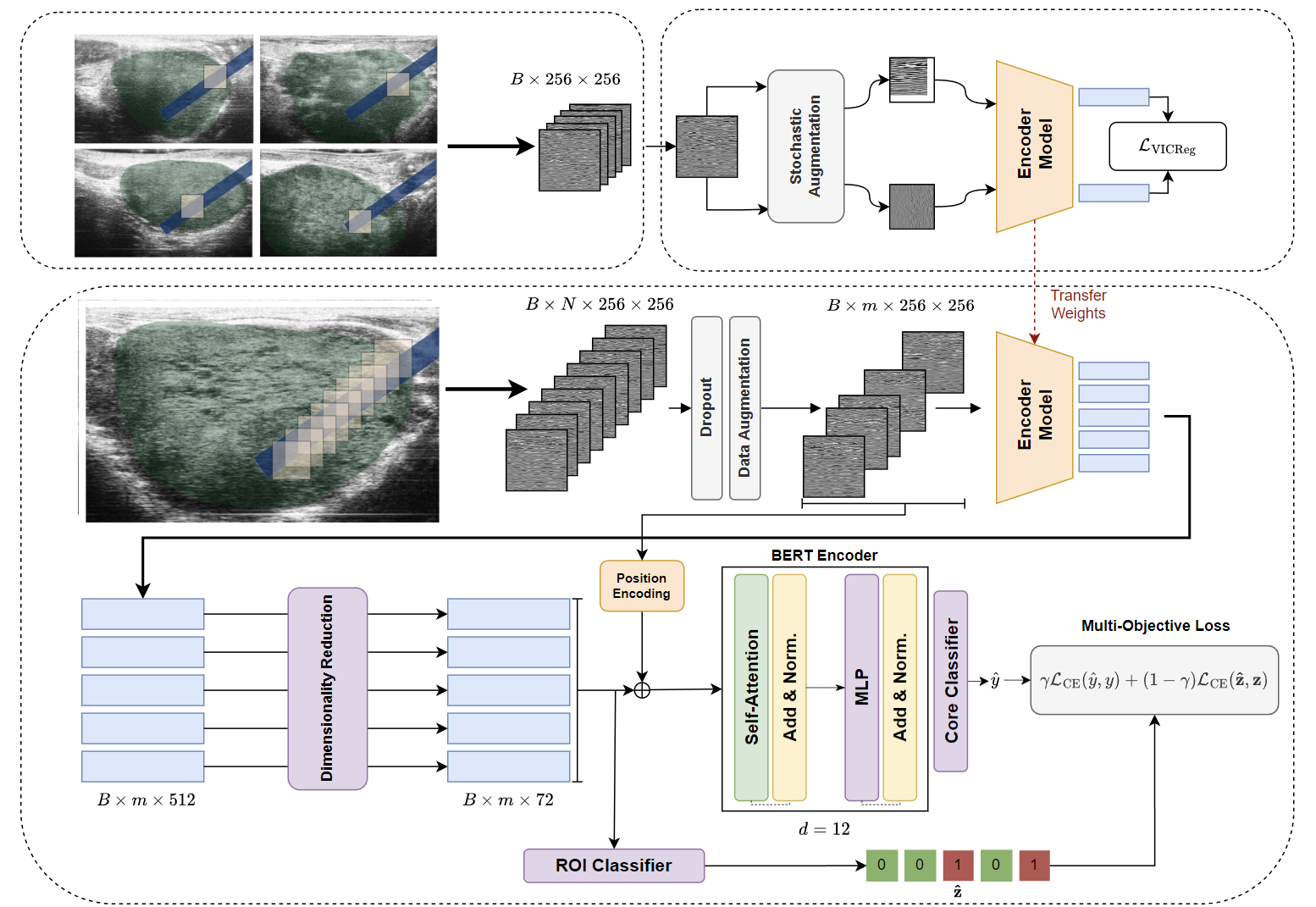}
    \vspace*{1.5mm}
    \caption{Multi-scale classification of prostate cancer across the whole biopsy core using BERT and multi-objective learning.}
    \label{fig:mo_trusformer}
\end{figure}

\noindent We pre-train each model described above using VICReg, then finetune them by loading the self-supervised weights and training the model with an additional MLP classifier attached. We train the model to detect cancer in individual ROIs, then match each ROI to their corresponding core. We then aggregate the predictions made for each patch in the core and compute their average to produce the final output of the core. 

\noindent We then take the 2 best models with the best performance and use them as a backbone to extract feature representations of all the ROIs in each core. We first project each feature representation to a $72\times1$ vector and train a 12-layer BERT\cite{devlin2018bert} classifier to produce a prediction for the entire core given this sequence of features. The weights of the model are updated using cross-entropy loss. 

\subsubsection{Multi-objective learning}
\noindent In addition to BERT sequence classification, we re-train the MLP layer used in Section 2.3 to assign a prediction to each ROI in the core. We then use those predictions as input to a second cross-entropy loss function. The final multi-objective loss can be summarized as follows:
\begin{align*}
    \mathcal{L}_\text{MO} &= \gamma \mathcal{L}_\text{CE} (\hat{y}, y) + (1-\gamma) \mathcal{L}_\text{CE} (\mathbf{\hat{z}}, \mathbf{z}),
\end{align*}
where $\mathcal{L}_\text{CE}$ designates the cross-entropy loss function, $\hat{y}$ is a core-wise prediction, $y$ is the label given to the entire core, $\mathbf{\hat{z}}$ is a vector containing predictions for each patch, and $\mathbf{z}$ is a vector containing the labels for all the patches (all of which are equal to $y$). $0 \leq \gamma \leq 1$ is a tunable hyperparameter that is used to weigh each component of the multi-objective loss. This workflow is illustrated in full in Figure~\ref{fig:mo_trusformer}.

\section{Results}
We evaluate both ROI-scale and multi-scale methods using nested $k$-fold cross validation, which divides the dataset into 5 folds each with an equal number of patients from each center. Each fold is used for testing once, with the remaining folds used for training and validation respectively. This ensures we obtain realistic estimates of the model's performance by evaluating on a wider variety of testing data. We compute the AUROC, sensitivity, and specificity of each model on each of the 5 test folds and average their performance. These results are shown in Table~\ref{tab:results_perform_table}.

\begin{table}[t]
    \centering
    \begin{tabular}{lccccc}
        \toprule
        \bf Backbone & \bf Finetuning & \bf AUROC & \bf Bal. Accuracy & \bf Sensitivity & \bf Specificity \\
        \midrule
        \em ROI-scale methods: \\
        ResNet18  & Linear & 76.1 $\pm$ 3.48 & 69.0 $\pm$ 2.51 & 65.3 $\pm$ 5.71 & \bf 73.6 $\pm$ 3.88 \\
        ViT  & Linear &66.8 $\pm$ 2.66 & 62.8 $\pm$ 2.91 & 69.1 $\pm$ 3.87 & 54.6 $\pm$ 2.09 \\
        CCT  & Linear & 74.1 $\pm$ 3.93 & 68.1 $\pm$ 2.80 & 70.6 $\pm$ 7.04 & 65.5 $\pm$  4.46\\ 
        PvT  & Linear &73.5 ± 3.70 & 67.4 $\pm$ 2.82 & 71.4 ± 9.56 & 63.4 $\pm$ 4.63 \\
        \midrule
        \em Multi-scale methods: \\
        ResNet18  & BERT & 76.6 $\pm$ 2.20 & 64.4 $\pm$ 1.50 & 63.4 $\pm$ 23.6 & 65.4 $\pm$ 25.9\\
        ResNet18  & BERT + MO$^*$ & \bf 77.9 $\pm$ 2.53 & \bf 71.1 $\pm$ 6.02 & \bf 75.9 $\pm$ 11.7 & 66.3 $\pm$ 19.7 \\
        CCT& BERT& 71.2 $\pm$ 3.79 & 61.5 $\pm$ 3.36 & 66.5 $\pm$ 27.1 & 56.5 $\pm$ 22.4\\
        CCT& BERT + MO$^*$ & 71.6 $\pm$ 3.01 & 63.4 $\pm$ 3.21 & 53.3 $\pm$ 13.9 & 73.4 $\pm$ 8.98\\
        \midrule
        \bottomrule
    \end{tabular}
    \vspace*{1.5mm}
    \caption{Comparing the performance of various ROI-based and core-based methods for prostate cancer detection}
    \label{tab:results_perform_table}
\end{table}

\noindent Among our ROI-level models, all 3 vision transformer variants fall short of exceeding the baseline ResNet model in overall performance, with a difference of $-2\%$ and $-0.9\%$ in AUROC and Balanced Accuracy for the CCT model. The standard ViT model drastically fails to keep up with the other ROI-scale models, with an AUROC of $66.8\%$, and an accuracy of $62.8\%$. Interestingly, the ResNet model obtains the lowest sensitivity among all ROI models, at $65.3\%$, lower than that of ViT, CCT, and PvT. CCT and PvT obtain the highest sensitivities at $70.6\%$ and $71.4\%$ respectively. This suggests that Transformer-based models are more likely to output a cancer prediction than not, whereas the ResNet baseline is more conservative, prioritizing negative predictions. 

\noindent Our MIL models show an improvement over single-instance ROI baselines, especially when using a ResNet18 backbone, with an AUROC exceeding that of any previous ROI baseline ($+0.5\%$ over ROI-scale ResNet18, $+2.5\%$ over CCT). Although the balanced accuracy of the MIL architecture is significantly smaller than that of its ROI-scale counterparts, choosing a threshold different than 0.5 at the output layer results in increased performance more in line with the AUROC. On the other hand, the MIL architecture with a CCT backbone fails to perform adequately, with a $5.1\%$ decrease in AUROC when compared to the baseline ResNet18 and a $3.1\%$ decrease when compared to the CCT model. We believe this shows that in the context of ultrasound-based PCa detection on a small dataset, convolutional features are more robust than Transformer embeddings. This would explain the performance of the ROI-scale models, as the 2 highest performing models use convolutions as part of their feature extraction. 

\noindent Finally, MIL models trained with multi-objective loss outperform the ones trained with simple cross-entropy, with a $1.3\%$ improvement in AUROC for the ResNet18+BERT architecture, and a modest $0.4\%$ improvement for CCT+BERT. As such, the multi-objective ResNet18+BERT obtains the highest performance metrics across all models, with an AUROC of $77.9\%$, a balanced accuracy of $71.1\%$, and a sensitivty of $75.9\%$, a significant improvement over all baselines. The CCT+BERT model continues to perform poorly even with multi-objective loss, failing to exceed its single-instance ROI counterpart ($-2.5\%$ AUROC, $-4.7\%$ Bal. Acc.). Hence, while CCT models perform reasonably well on ROI-scale PCa detection, they do not translate as well to a multi-scale setting. These results suggest that ResNet18 models are the most effective feature extractors for our micro-ultrasound dataset.

\footnotetext[1]{\underline{M}ulti-\underline{O}bjective loss function}

\section{Conclusion}
We conclude that given a small dataset of prostate ultrasounds such as ours, feature representations learned by a Transformer backbone are insufficient to exceed convolutional baselines in performance. One possible reason for this could be the better parameter efficiency afforded by convolutional layers, which mitigates overfitting. It is also possible that convolutions are better suited for modeling features extracted from prostate ultrasounds, resulting in stronger performance. Finally, we find that using multi-objective learning to combine both ROI and core loss functions improves performance sufficiently to beat both the ROI-scale baseline and the multi-scale state of the art method for PCa detection from ultrasound.

\acknowledgments
We thank the Natural Sciences and Engineering Research Council of Canada (NSERC) and the Canadian Institutes of Health Research (CIHR) for supporting our work. Parvin Mousavi is supported by the CIFAR AI Chair and the Vector Institute. Brian Wodlinger is Vice President of Clinical and Engineering at Exact Imaging. All patient data was used with informed consent and approval of institutional ethics boards.
 

\bibliographystyle{spiebib}

\begin{thebibliography}{10}

\bibitem{ahmed2017diagnostic}
Ahmed, H.~U., Bosaily, A. E.-S., Brown, L.~C., Gabe, R., Kaplan, R., Parmar, M.~K., Collaco-Moraes, Y., Ward, K., Hindley, R.~G., Freeman, A., et~al., ``Diagnostic accuracy of multi-parametric mri and trus biopsy in prostate cancer (promis): a paired validating confirmatory study,'' {\em The Lancet}~{\bf 389}(10071),  815--822 (2017).

\bibitem{wilson2022self}
Wilson, P.~F., Gilany, M., Jamzad, A., Fooladgar, F., To, M. N.~N., Wodlinger, B., Abolmaesumi, P., and Mousavi, P., ``Self-supervised learning with limited labeled data for prostate cancer detection in high frequency ultrasound,'' {\em arXiv preprint arXiv:2211.00527}  (2022).

\bibitem{gilany2023trusformer}
Gilany, M., Wilson, P., Perera-Ortega, A., Jamzad, A., To, M. N.~N., Fooladgar, F., Wodlinger, B., Abolmaesumi, P., and Mousavi, P., ``Trusformer: improving prostate cancer detection from micro-ultrasound using attention and self-supervision,'' {\em International Journal of Computer Assisted Radiology and Surgery} ,  1--8 (2023).

\bibitem{rohrbach2018high}
Rohrbach, D., Wodlinger, B., Wen, J., Mamou, J., and Feleppa, E., ``High-frequency quantitative ultrasound for imaging prostate cancer using a novel micro-ultrasound scanner,'' {\em Ultrasound in medicine \& biology}~{\bf 44}(7),  1341--1354 (2018).

\bibitem{bardes2021vicreg}
Bardes, A., Ponce, J., and LeCun, Y., ``Vicreg: Variance-invariance-covariance regularization for self-supervised learning,'' {\em arXiv preprint arXiv:2105.04906}  (2021).

\bibitem{dosovitskiy2020image}
Dosovitskiy, A., Beyer, L., Kolesnikov, A., Weissenborn, D., Zhai, X., Unterthiner, T., Dehghani, M., Minderer, M., Heigold, G., Gelly, S., et~al., ``An image is worth 16x16 words: Transformers for image recognition at scale,'' {\em arXiv preprint arXiv:2010.11929}  (2020).

\bibitem{hassani2021escaping}
Hassani, A., Walton, S., Shah, N., Abuduweili, A., Li, J., and Shi, H., ``Escaping the big data paradigm with compact transformers,'' {\em arXiv preprint arXiv:2104.05704}  (2021).

\bibitem{wang2021pyramid}
Wang, W., Xie, E., Li, X., Fan, D.-P., Song, K., Liang, D., Lu, T., Luo, P., and Shao, L., ``Pyramid vision transformer: A versatile backbone for dense prediction without convolutions,'' in [{\em Proceedings of the IEEE/CVF international conference on computer vision}{\nolinebreak\hspace{0.1em}]},   568--578 (2021).

\bibitem{he2016deep}
He, K., Zhang, X., Ren, S., and Sun, J., ``Deep residual learning for image recognition,'' in [{\em Proceedings of the IEEE conference on computer vision and pattern recognition}{\nolinebreak\hspace{0.1em}]},   770--778 (2016).

\bibitem{devlin2018bert}
Devlin, J., Chang, M.-W., Lee, K., and Toutanova, K., ``Bert: Pre-training of deep bidirectional transformers for language understanding,'' {\em arXiv preprint arXiv:1810.04805}  (2018).

\end{thebibliography}

\end{document}